\documentclass[aps,twocolumn,showpacs,superscriptaddress,amssymb]{revtex4}
\date{\today}
\usepackage{graphicx}
\begin{document}

\title{Modern shell-model diagonalizations with realistic \emph{NN} forces}

\author{C. Qi}
\affiliation{School of Physics and MOE
Laboratory of Heavy Ion Physics, Peking University, Beijing 100871,
China}
\author{F.R. Xu}
\email{frxu@pku.edu.cn} \affiliation{School of Physics and MOE
Laboratory of Heavy Ion Physics, Peking University, Beijing 100871,
China} \affiliation{Institute of Theoretical Physics, Chinese
Academy of Sciences, Beijing 100080, China} \affiliation{Center for
Theoretical Nuclear Physics, National Laboratory for Heavy Ion
Physics, Lanzhou 730000, China}

\begin{abstract}
The spectral and statistical properties of nuclei $^{46}$V and
$^{48}$Cr are studied in the framework of nuclear shell model. A
microscopical effective Hamiltonian derived from the CD-Bonn
\textit{NN} potential is employed. The calculations are preliminary
results of a newly developed parallel shell-model code. Calculations
with the realistic \textit{NN} interaction reproduce well the
ground-state rotational band and backbending phenomenon in
$^{48}$Cr. A sudden alignment of all particles is seen above the
backbend. The nearest-neighbour spacing distribution in $^{46}$V is
presented and compared with the predictions of random-matrix theory.
Significant deviation from the single Gaussian-orthogonal-ensemble
distribution is obtained, indicating that the isospin-symmetry
breaking is not so strong in the nucleus.
\end{abstract}

\maketitle

The nuclear shell model has been considered as the most fundamental
tool for the investigation of nuclear structures and electromagnetic
and weak transitions. It provides the microscopic basis to study the
nature of nuclear states, magnetic and quadrupole moments, $\beta$
and $\gamma$ transitions and nuclear reactions. The simpliest
version of the shell model, the so-called independent particle
model, has been proposed more than fifty years ago and gives
successful descriptions of the nuclear magic numbers and
ground-state angular momenta with the introduction of strong
spin-orbit coupling interaction \cite{Talmi03}. Soon, the community
realized the importance of introducing configuration-mixing in
explaining the spectroscopic properties of nuclei (i.e., $^{19}$F)
with singular ground-state spins \cite{Ta54,Elliott55}. The
configuration-mixing concept has been considered as the essence of
modern shell model calculations \cite{Brown01,Dean04,Caurier05}.

Now the shell model has been well understood in the language of
perturbation theory. The basis to study the nuclear many-fermion
systems is a non-relativistic Schr\"odinger equation which has
infinite freedom and insolvable. As done in atomic many-body
studies, the starting point is to reduce the Schr\"odinger equation
to finite freedom with the help of Feshbach operators
\cite{Haxton,atom}. The Schr\"odinger equation can be written in the
form
\begin{equation}\label{feshbach}
\left(
\begin{array}{cc}
PHP & PHQ \\
QHP & QHQ \\
\end{array}
\right)\left(
\begin{array}{c}
P\Psi \\
Q\Psi \\
\end{array}
\right)=E\left(
\begin{array}{c}
P\Psi \\
Q\Psi \\
\end{array}
\right),
\end{equation}
where $H$ is the original Hamiltonian, $E$ the eigen energies and
$P$ and $Q$ projection operators defining the model space and the
excluded orthogonal space. From Eq. (\ref{feshbach}) we can derive
an model-space dependent effective Hamiltonian which is given
 by \cite{bloch}
\begin{equation}\label{eff}
H_{\text{eff}}=P\{H+H\frac{1}{E-QH}QH\}P.
\end{equation}
The original eigen energies and wave functions can be constructed by
diagonalizing this equivalent Hamiltonian. The effective Hamiltonian
in Eq. (\ref{eff}) is energy dependent and can be solved
self-consistently \cite{Haxton} or perturbatively \cite{atom}. In
the latter case the effective Hamiltonian can be written as
\begin{equation}
H_{\text{eff}}=PH_0P+PV\Omega P,
\end{equation}
where $H_0$ is the unperturbed Hamiltonian, $\Omega$ the wave
function operator and $V$ the nuclear force. The last term is
referred to as the effective residual interaction with
$v_{\text{eff}}=PV\Omega P$. In the shell model context, the single
particle Hamiltonian is evaluated from the experimental observations
of the single particle outside the assumed core. The effective
interaction $v_{\text{eff}}$ is usually treated perturbatively up to
two-body level and expressed as two-body matrix elements in harmonic
oscillator (HO) basis. The shell structure of nuclear systems
recognized by Mayer {\it et al.} \cite{Talmi03} makes it possible to
constraint calculations in one or several major shells defined by
the projection operator $P$ which can embed the dominant component
of the the original wave function.

Although calculations can be restricted in a single major shell, the
corresponding Hilbert space dimension is still very large. The
\emph{fp} shell (with M-scheme dimension around $10^9$) presents the
state-of-the-art shell-model performance in the past decade.
Diagonalizations for nuclei in the intermediate \emph{fp} model
space (e.g., $^{56}$Ni), however, are still painstaking
\cite{Horoi06}. Hence, the development and optimization of
shell-model diagonalization algorithm has been a longstanding work
\cite{oak,Oxbash,Whitehead,antoine,Toivanen,Dean04,Dusm,tokyo,redstick,Horoi06}.
The first implementation of efficient shell-model diagonalization is
the Oak-Ridge Rochester code \cite{oak} using the \emph{j-j} coupled
scheme \cite{Talmi03}. Another revolutionary development attributes
to the Glasgow group \cite{Whitehead}. The algorithm implemented by
Whitehead {\it et al.} \cite{Whitehead} introduced the M-scheme
representation and the Lanczos diagonalization method. The M-scheme
is compatible with basic bit operations in modern computer and
avoids the time-consuming calculations of coefficients of fractional
parentage (CFP) in \emph{j-j} scheme. The Lanczos algorithm uses an
iterative procedure and is suitable for giant matrix
diagonalizations \cite{Dean04,antoine}. The powerful shell-model
code ANTOINE of the Strasbourg-Madrid group succeeded the idea of
the Glasgow algorithm \cite{antoine}.

In M-scheme, only $J_z$ and $T_z$ are good quantum numbers, leading
to a maximal dimension of the generated Hamiltonian matrix. The
famous OXBASH code and its MSU version \cite{Oxbash} follows a
hybrid algorithm between the M-scheme and the \emph{j-j} coupled
scheme. The symmetry of $J$ ($T$) is restored through a projection
procedure. The dimension of the matrix can be significantly reduced
through this symmetry restoration. The algorithm implemented in
OXBASH makes it very efficient to calculate many converged vectors
for a given angular momentum. However, the projection process is
very time-consuming and has precision problems.

\begin{figure}
\includegraphics[scale=0.37]{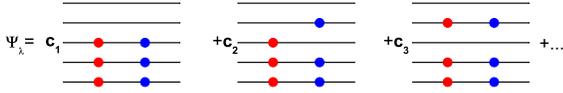}\\
\caption{Schematic picture for shell-model wave function as an
expansion of partitions. The lines indicate single-particle orbits
in the HO basis.}\label{part}
\end{figure}

In this paper we report on preliminary results calculated with a
newly developed parallel shell-model code. It follows the hybrid
algorithm of OXBASH and embeds a projection procedure to generate
basis with good $J$ quantum number. The program is written in
Fortran 90 format which is more efficient and is more suitable for
high performance computing than the outmoded FORTRAN 77 standard
\cite{fort90}. In Fortran 90 language, the precision of data objects
can be easily defined, while super-large matrices and arrays can be
handled. The code is implanted on the 64-bit Beowulf cluster of PKU
computer center. Parallel algorithms are realized with the help of
Message Passing Interface (MPI) \cite{mpi}.

\begin{figure}
\includegraphics[scale=0.39]{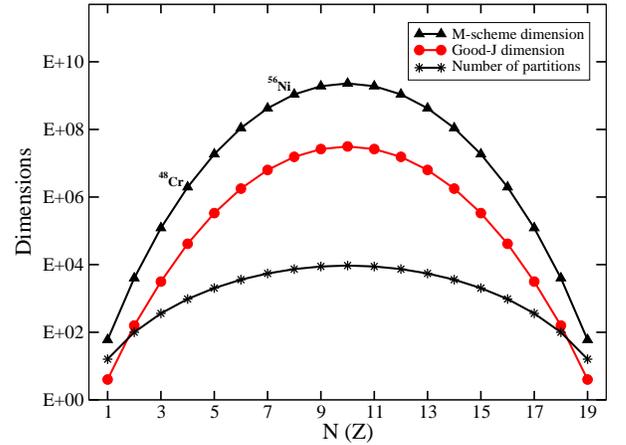}\\
\caption{M-scheme and $j-j$ coupled scheme dimensions for $0^{+}$
states in \emph{fp} shell \emph{N=Z} nuclei. Stars denote the
numbers of partitions for each state.}\label{dim}
\end{figure}

The starting point of the calculation is to generate a set of bases
with good magnetic quantum number in M-scheme. Neutrons and protons
are blocked into two spaces.  The bases are classified according to
the distributions of particles in the single-particle orbits
(referred to as partitions). Schematic figure for shell model
wave-function as an expansion of partitions is plotted in Fig.
\ref{part}. Since the angular-momentum projection operator can only
change the magnetic quantum number, projected vectors in different
partitions are orthogonal, ensuring that projections can be done for
each partition separately. In Fig. \ref{dim} we plotted M-scheme
dimensions, \emph{j-j} scheme dimensions and number of partitions
for $0^+$ states in \emph{fp}-shell $N=Z$ nuclei as a function of
neutron (proton) numbers.

Vectors with good angular momentum are expanded in M-scheme bases
for each partition, which can be given as
\begin{equation}
|\Psi^{J}_i\rangle=\sum_{m\leq i}
\mathcal{M}_{im}P^{J}|\alpha_m\rangle,
\end{equation}
where $P^J$ is the projection operator and $|\alpha\rangle$ a set of
specially chosen M-scheme bases. $\mathcal{M}$ is a lower triangle
matrix. Vectors with good angular momentum are the orthonormal and
satisfy,
\begin{eqnarray}
\nonumber
\langle\Psi^{J}_i|\Psi^{J}_j\rangle&=&\langle\Psi^{J}_i|\sum_{m\leq
j} \mathcal{M}_{jm}P^{J}|\alpha_m\rangle\\
\nonumber &=&\sum_{m\leq j} \mathcal{M}_{jm}\langle
\Psi^{J}_i|\alpha_m\rangle\\
&=&\delta_{ij}.
\end{eqnarray}
Elements of $\mathcal{M}$ can be obtained by inverting the matrix
$\langle \Psi^{J}_i|\alpha_m\rangle$. Since the projection operator
satisfies, $P^J\cdot P^J=P^J$, the $n$-th vector can be obtained
through,
\begin{equation}
|Q_n\rangle=|\alpha_n\rangle-\sum_{i<n}\langle O_i|\alpha_n\rangle
\sum_{j\leq i}\mathcal{M}_{ij}|\alpha_j\rangle,
\end{equation}
and
\begin{equation}
|\Psi^J_n\rangle=P^J|Q_n\rangle(\langle Q_n|P^J|Q_n\rangle)^{-1/2}.
\end{equation}
In the meanwhile, matrix elements in the $n$-th row of the
$\mathcal{M}$ is calculated. This projection algorithm is very
efficient since calculating operations and the number of
coefficients needed in the calculation are comparatively small and
can be stored in the computer memory. Projections for various
partitions can be done independently and can run parallelly with
very high scalability.

\begin{figure}
\includegraphics[scale=0.35]{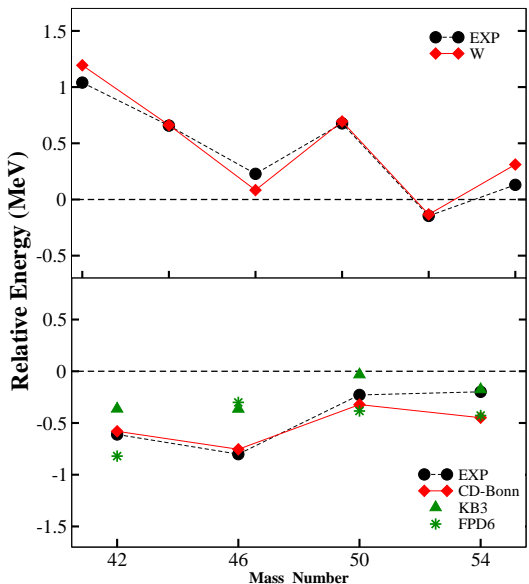}\\
\caption{\label{sdfp}Relative energies of first $T=1$ states
($J^{\pi}=0^+$) and first $T=0$ states for odd-odd $N=Z$ nuclei
($T_z=0$) in the \emph{sd} and $0f_{7/2}$ shells. Experimental data
are taken from Ref. \cite{rad}. Calculations for \textit{sd} and
$0f_{7/2}$ shells are done with the W \cite{USD} and CD-Bonn
\cite{interaction}, KB3 \cite{kb3} and FPD6 \cite{fpd6}
interactions, respectively.}
\end{figure}

The projection procedure shown above was suggested but not used in
OXBASH \cite{Oxbash} since it does not conserve $t_z$ values and is
unsuitable for projections in isospin space. The advantages of
isospin space lie in that only the smaller set of $(J,T)$ vectors
satisfying $J=J_M$ and $T=T_z$ are projected. The number of non-zero
Hamiltonian matrix elements is also smaller. The restriction of
$J=J_M$ is valid since the Hamiltonian is invariant under rotations
in angular momentum space. However, nuclear states with $T>T_z$ do
exist and can exist at very low excitation energies. Examples in
odd-odd and even-even nuclei are shown in Fig. \ref{sdfp} and Table
\ref{be2}, respectively. To give the full level structure, besides
the generation of the smaller $(J,T)$ set with $T=T_z$, projections
in isospin space have to be imposed to the maximum of the $T$ value.

\begin{table}
\centering \caption{Examples of states in even-even nuclei that have
$T>|T_z|$. Calculations are done using the W interaction
\cite{USD}.}\label{be2}
\begin{ruledtabular}
\begin{tabular}{llcc}
Nuclei&$J^{\pi}$&Cal.~(MeV)&Exp.~(MeV)\\
\hline
$^{20}$Ne&$1^+_1$&11.20&11.26\\
$^{24}$Mg&$1^{+}_3$&9.99&9.97\\
$^{28}$Si&$1^+_3$&10.81&10.72\\
$^{32}$S&$1^+_2$&7.06&7.00\\
$^{36}$Ar&$2^+_3$&6.65&6.61\\
$^{36}$Ar&$3^+_2$&7.46&7.34\\
$^{36}$Ar&$1^+_3$&8.19&8.13
\end{tabular}
\end{ruledtabular}
\end{table}

Since the nuclear force $V$ is a scalar product and conserves spin
and parity, the effective interaction can be simply expressed as
\begin{equation}
v_{\text{eff}}=\sum_{J}\sum_{\alpha\leq
\beta;\gamma\leq\delta}\langle\alpha\beta;J|V|\gamma\delta;J\rangle,
\end{equation}
where $\alpha$ denotes the single-particle orbit. If isospin is
conserved in the Hamiltonian, the two-body matrix elements can be
further simplified by adding isospin quantum numbers to the
proton-proton and neutron-neutron ($T=1$) and proton-neutron ($T=0$
and 1) interactions, which are restricted by $J+T$=odd for
$\alpha=\beta$ and/or $\gamma=\delta$. The Hamiltonian matrix
elements are given as,
\begin{equation}\label{hm}
\langle \Psi_i^J|H|\Psi_j^J\rangle=\sum_n
\mathcal{M}_{jn}\langle\Psi_i^J|H|\alpha\rangle.
\end{equation}
In practical calculations, the matrix
$\langle\Psi_i^J|H|\alpha\rangle$ are calculated first and store in
memory. The total matrix elements are generated by scanning and
timing the $\mathcal{M}$ matrix. The calculation of Hamiltonian
matrix elements can be separated by partition and done parallelly.
However, the efficiency is not optimal when dimensions are not very
giant. Another very efficient way is to separate the above two steps
into two or several computing processes.

The Hamiltonian matrix is diagonalized using the Lanczos procedure
\cite{Dean04,Whitehead}. The most time-consuming part is the
iterative generating of Lanczos vectors,
\begin{equation}
\beta_{n+1}\rangle=H|\beta_n\rangle,
\end{equation}
where $|\beta_0\rangle$ is a specially chosen basis in \emph{j-j}
scheme. In the present code, the Hamiltonian matrix elements are
scattered into available processes using the MPI paradigm
\cite{mpi}, in which the matrix-array production can be done
parallelly and be reduced to give out the new Lanczos vectors.

The maximum capacity of the computer code has not been tested, which
can be restricted by the Input/Output operation rate of the
computing system. In this work we present calculations for nuclei
$^{46}$V and $^{48}$Cr. The largest \emph{j-j} scheme dimension is
about $2.5 \times 10^5$, for which the scalability of the
diagonalization using up to 30 processors is still optimal.

\begin{figure}
\includegraphics[scale=0.36]{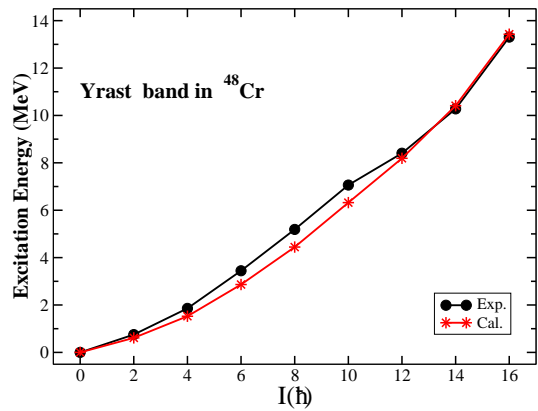}\\
\caption{Calculated and experimental yrast band in
$^{48}$Cr.}\label{cr48}
\end{figure}

\begin{figure}
\includegraphics[scale=0.36]{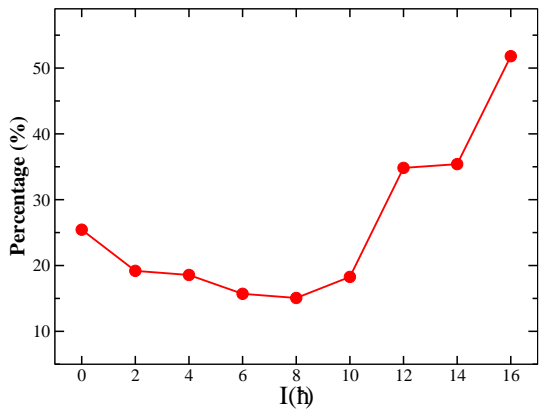}\\
\caption{The percentage of pure $0f_{7/2}^8$ configuration in the
wave functions of yrast states in $^{48}$Cr.}\label{fig5}
\end{figure}

In the past decade, the medium-mass nucleus $^{48}$Cr, which has a
half-filled $0f_{7/2}$ shell of protons and neutrons, has been under
intensive study (see e.g. Ref. \cite{Juodagalvis00} and reference
therein). A well-deformed ground state appears in $^{48}$Cr,
together with many macroscopic phenomena, like rotation, backbend
and triaxial deformation \cite{Juodagalvis00,Caurier95}. Previous
shell-model study by Caurier {\it et al.} using a modified Kuo-Brown
interaction reproduced the yrast band and the backbending phenomenon
\cite{Caurier95}.

In the present work, We used an isospin-nonconserving effective
Hamiltonian to study the structure of $^{48}$Cr in full \emph{fp}
shell. The effective interaction is derived microscopically from a
high-precision charge-dependent Bonn potential \cite{interaction}.
Single-particle energies are taken from experimental observations of
odd neutron and odd proton with respect to the core of $^{40}$Ca, in
which the Thomas-Ehrman shift of proton $1p$ orbits can also be
taken into account approximately. Excellent agreements with
experiments is seen till the band terminated state with $I=16^+$.
Calculations characterize the backbend appeared at $I=12^+$.
Detailed analyse of wave functions shows that, for all states in
Fig. \ref{cr48}, the average numbers of particles in $0f_{7/2}$
orbit (or $0f_{7/2}$ orbital occupancy) are nearly constant (around
six). Above the backbend, particles are suddenly aligned, which
manifest itself in that pure $0f_{7/2}$ configuration significately
increases for states with $I\geq12$. The dispersion of
configurations at low spins contributes to the deformation.

The shell model provides the most promising foundation to
investigate the complexity and chaotic behavior in nuclear many-body
systems, which is the best test-ground for the study of quantum
chaos. Comprehensive reviews on the topic can be find, e.g., in
Refs. \cite{Guhr98,Zelevinsky96}. Now it is believed that the level
fluctuation properties (i.e., the nearest-neighbour spacing
distribution (NNS)) of quantum systems with time-reversal symmetry
whose classical analogs are chaotic can well described by the
Gaussian orthogonal ensemble (GOE) of random-matrix theory (RMT),
whereas quantum analogs of classically integrable systems follow the
Poisson statistics. Particularly interesting is the statistical
properties of systems with one additional approximate symmetry. If
the symmetry is exact, the NNS distribution follows the
superposition of two independent GOE (2-GOE) \cite{Guhr98}. The
complete absence of the symmetry in the distribution, however, is
expected even with a small breaking of the symmetry
\cite{Mitchell88}. A vivid example has been shown in the spectral
statistics of acoustic resonances in rectangular quartz blocks, in
which a transition from 2-GOE to GOE is seen with the gradual
breaking of the point-group symmetry in the system
\cite{Ellegaard96}.

\begin{figure}
\includegraphics[scale=0.36]{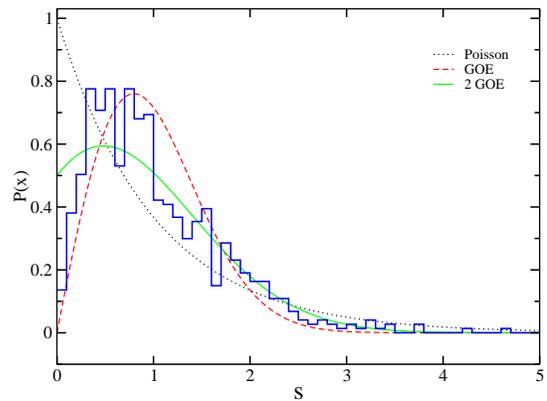}\\
\caption{The NNS distribution for low-lying levels in $^{46}$V with
$J<15$.}\label{v46all}
\end{figure}

\begin{figure}
\includegraphics[scale=0.36]{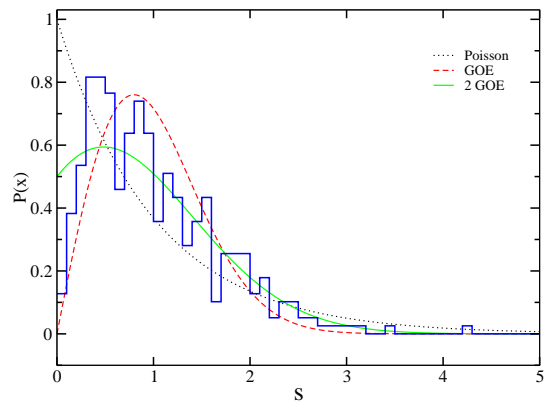}\\
\caption{Same as Fig. \ref{v46all} but only for states with even
spins.}\label{v46e}
\end{figure}

Systematical works have been done to study the chaotic behaviors of
 \emph{sd}-shell nuclei, both experimentally and theoretically
\cite{Zelevinsky96}. The chaotic behavior in \emph{sd} shell nuclei
has been well described by shell model calculations using the
empirical \emph{W} interaction \cite{USD}. However, controversy
still exists concerning the role played by isospin symmetry breaking
on the statical properties of NNS \cite{Mitchell88}. In nucleus
$^{26}$Al, a system with $5p-5n$ above the core, the NNS
distribution for low-lying states generated by empirical
isospin-nonconserving Hamiltonian shows a strong isospin dependence,
indicating that the isospin symmetry breaking is small, whereas a
strong isospin symmetry breaking effect is expected from the analyse
of available observations \cite{Mitchell88}. In the \textit{sd}
shell, strong Coulomb influence exists due to the large
Thomas-Ehrman shift of proton $1s_{1/2}$ orbit. This can lead to
large uncertainties \cite{Mitchell88}. One of the notable feature of
the \textit{fp} shell is that the isospin symmetry breaking is
relatively smaller \cite{interaction}. It would be very helpful to
see the statistical behavior in the shell and the isospin symmetry
breaking influence. Deviation from the GOE distribution has been
noticed in the NNS distribution for the low-lying states of Ca
isotopes \cite{Caurier96}.

We evaluate the isospin symmetry breaking effect in the NNS
distribution of $^{46}$V using the isospin-nonconserving force
\cite{interaction} discussed above. The lowest 150 positive-parity
states for each spin between $I=0$ and 16 ($0f_{7/2}$ band
termination is at $I=15$) were calculated. The 2550 states can be
generated on a single processor within an hour. The energy levels
for each spin are unfolded through \cite{Caurier96}
\begin{equation}
\bar{N}(E)=\int_0^E\bar{\rho}(E')dE'+N_0,
\end{equation}
where $\bar{\rho}(E)$ is the mean level density which is assumed to
be a exponential form,
\begin{equation}
\bar{\rho}(E)=1/T\cdot\text{exp}[(E-E_0)/T].
\end{equation}
The distribution of NNS, $P(s)$, is obtained by accumulating the
number of spacings of $s_i=\bar{N}(E_{i+1})-\bar{N}(E_i)$ within
$(s,s+\Delta s)$. Results are plotted in Fig. \ref{v46all} and
\ref{v46e}.

Our calculations are carried out in proton/neutron space and is
irrespective of isospin. The isospin symmetry breaking in the
effective Hamiltonian has three origins: the Coulomb shifts of
proton $1p$ single-particle orbits ($\sim200$ keV), the charge
dependence in strong force and two-proton Coulomb matrix elements.
If isospin symmetry is exact, the distribution should follow the
2-GOE distribution, while a transition to the single GOE
distribution is expected with the gradual breaking of isospin
symmetry. Our calculations show that, however, the deviation from
the 2-GOE distribution is relatively negligible. Other statistical
properties, like the Brody parameter and the spectral rigidity, can
provide more deep insight on this problem.

In conclusion, a new high-performance parallel shell-model code has
been developed. With the code we calculated the spectral and
statistical properties of nuclei $^{46}$V and $^{48}$Cr with a
microscopical effective Hamiltonian derived from the CD-Bonn
\textit{NN} potential. Calculations with the realistic \textit{NN}
interaction reproduce well the yrast band and backbending phenomenon
in $^{48}$Cr. A sudden increasement of pure $0f_{7/2}$ configuration
is seen above the backbend. In the meanwhile, the average numbers of
particles in the yrast states remains nearly constant. The NNS
distribution for low-lying states in $^{46}$V is calculated and
compared with the predictions of random-matrix theory. The
isospin-nonconserving Hamiltonian is used to evaluate the isospin
symmetry breaking effect on the NNS distribution. Significant
deviation from the GOE distribution is obtained, indicating that the
isospin-symmetry breaking is not so strong in the nucleus.

\section*{Acknowledgement}
We thank Fan Chun, Wu Zhe-ying and Pei jun-chen for useful
discussions on parallel computing. This work has been supported by
the Natural Science Foundations of China under Grant Nos. 10525520
and 10475002, the Key Grant Project (Grant No. 305001) of Ministry
of Education of China. We also thank the PKU computer Center where
the parallel code has been implanted and numerical calculations have
been done.

\end{document}